\documentclass[aps,twocolumn,amsfont,amssymb,amsmath,superscriptaddress]{revtex4-2}
\usepackage[sort&compress]{natbib}
\usepackage{graphicx}
\usepackage[normalem]{ulem}
\usepackage{mathtools}
\usepackage{dcolumn}
\usepackage{bm}
\usepackage{hhline}
\usepackage[dvipsnames]{xcolor}
\usepackage{hyperref}
\usepackage{float}
\hypersetup{
    colorlinks,
    linkcolor={blue!70!black},
    citecolor={blue!70!black},
    urlcolor={blue!70!black}
}
\newcommand{\mvec}[1]{\mathbf{#1}}
\newcommand{\sgrad}{\nabla}

\newcommand{\sdiv}{\nabla\cdot}
\newcommand{\sdot}[2]{\mathbf{#1}\cdot\mathbf{#2}}
\newcommand{\lap}{\nabla^2}

\begin{document}
	\title{Coarsening of topological defects in 2D polar active matter}
	\author{Soumyadeep Mondal}
        \email{msoumyadeep@iisc.ac.in}
	\author{Pankaj Popli}
     \email{pankajpopli@iisc.ac.in}
	\author{Sumantra Sarkar}
        \email{sumantra@iisc.ac.in}
        
	\affiliation{Center for Condensed Matter Theory, Department of Physics, Indian Institute of Science, Bengaluru, Karnataka, India, 560012}
	\date{\today}
	\begin{abstract}
		We numerically study the coarsening of topological defects in 2D polar active matter and make several interesting observations and predictions. (i) The long time state is characterized by nonzero density of defects, in stark contrast to theoretical expectations. (ii) The kinetics of defect coarsening shows power law decay to steady state, as opposed to exponential decay in thermal equilibrium. (iii) Observations (i) and (ii) together suggest emergent screening of topological charges due to activity. (iv) Nontrivial defect coarsening in the active model leads to nontrivial steady state patterns. We investigate, characterize, and validate these patterns and discuss their biological significance. 
	\end{abstract}
	\maketitle
	
     Active materials consist of particles that stay inherently out of equilibrium by converting supplied energy into work~\cite{ramaswamy2010mechanics, marchetti2013hydrodynamics}. Unlike externally-driven non-equilibrium systems, the energy generation and dissipation in active materials happen at the level of individual entities. The interactions among these active particles result in rich, diverse collective behaviors spanning from microscopic scales to kilometers. For example, the collective movement of bacteria, the arrangement of cytoskeleton filaments within cells, and the flocking behavior of birds etc.~\cite{marchetti2013hydrodynamics}.

Topological defects are nontrivial configurations allowed by the symmetries of the system and correspond to the discontinuities in the order parameter field, which cannot be annealed out by any smooth deformations of the field. For instance, in 2D XY model, they manifest as point vortices or saddles~\cite{mermin1979topological,CL-book,pismen-book}. They are characterized by their \textit{topological charge}, which is the winding number of the singular order-parameter field at the defect core. In equilibrium, defects with opposite charges unbind above a nonzero critical temperature, $T_{KT}$, destroying quasi-long-range order~\cite{kosterlitz2018ordering}. In active systems, in contrast, topological defects unbind even at zero temperature~\cite{marchetti2013hydrodynamics}, which makes the homogeneous ordered state unstable to fluctuation above a length scale~\cite{simha2002hydrodynamic}. The dynamics of individual active defects also differ significantly from the passive cases. Whereas, passive defects move purely through diffusive motion~\cite{yurke1993coarsening}, active defects, depending on their symmetries, may also self-propel. For example, in 2D active nematic materials, +1/2 topological defects self-propel, leading to active turbulence~\cite{thampi2016active,alert2022active}. In 2D active polar materials, the rotational symmetries of $\pm$1 defects prevent any self-propulsion in general, but when the rotational symmetry is broken, self-propulsion is observed~\cite{husain2017emergent}. 
	
	Interactions between many topological defects determine the emergent macroscopic properties of active materials~\cite{decamp2015orientational,lemma2022active,chardac2021emergence,chardac2021topology,rana2022phase,saha2014clusters}. Understanding the dynamics of many defects has been of recent interests~\cite{shankar2018defect,shankar2019hydrodynamics,vafa2022defect}. In active nematic systems, the defects can lead to chaotic or polar organization of $\pm$1/2 charges~\cite{shankar2019hydrodynamics}. In 2D polar active systems, such as the actin cytoskeleton (ACS), the interaction and dynamics of multiple defects shape the macroscopic organization of the material, leading to nontrivial dynamic patterns~\cite{goswami2008nanoclusters, gowrishankar2010active,gowrishankar2016nonequilibrium}. In cells, such patterns in the ACS have been hypothesized to control the size~\cite{goswami2008nanoclusters,gowrishankar2010active} and the lifetime of clusters of membrane proteins~\cite{sarkar2023lifetime}. In experiments with polar filaments, it has been observed that coarsening of topological defects is a key driver of pattern formation~\cite{soares2011active,wollrab2019polarity}. However, the underlying mechanism of coarsening and the resultant patterns are poorly understood.  
 
 In this paper, we investigate the coarsening dynamics of topological defects in a model of 2D polar active matter~\cite{gowrishankar2010active} in the absence of thermal and active noise. We find that defects merge with each other until an activity dependent threshold density is reached. The threshold density decreases with decreasing activity, recovering the passive behavior at the zero activity limit, implying the screening of the long-range interaction between the topological defects. To investigate the origin of the screening, we study the kinetics of defect coalescence and identify a mechanism that is distinct from equilibrium coarsening. Finally, we validate our findings by comparing our predictions with live-cell measurements of ACS heterogeneity~\cite{xia2019nanoscale}.

	\paragraph*{Model} We study the following model of 2D polar active matter. 
\begin{eqnarray}
		\partial_t \mvec{p} &+& \lambda (\sdot{p}{\sgrad})\mvec{p} = K \lap{\mvec{p}} +  ( \alpha - \beta |\mvec{p}|^2)\mvec{p} + \zeta\sgrad{c} +  \mvec{f_{\mvec{p}}}~~~ \label{eqn:p}\\	
		\partial_t c &=& -\sdiv{(\mvec{j_a} + \mvec{j_d})}=-\sdiv{(v_0c\mvec{p} - D\sgrad c)}  \label{eqn:c}
	\end{eqnarray}
Eq.\ref{eqn:p} describes the evolution of the 2D polar field $\mvec{p} \equiv (p_x,p_y)$, coupled to a concentration field $c$. Eq.\ref{eqn:c} is a continuity equation ensuring the conservation of the total number of active particles transported through active ($\mvec{j_a}$) and diffusive ($\mvec{j_d}$) currents. The first two terms on the right hand side of Eq.~\ref{eqn:p} arises from the free energy functional of the 2D XY model~\cite{gowrishankar2010active}. The third, active, term couples $\mvec{p}(\mvec{r},t)$ to $c(\mvec{r},t)$ through the activity parameter (called contractility), $\zeta$.  In Eq.\ref{eqn:c}, the active term couples the two fields through the self-propulsion speed $v_0$, which we take to be 1. This model is useful in studying many biological systems, such as the actin cytoskeleton, where the concentration and the orientation of the active constituents are coupled to each other~\cite{gowrishankar2010active}. We simplified Eq.~\ref{eqn:p} further by taking advection coefficient $\lambda = 0$ and noise $\mvec{f_p}=\mvec{0}$, which does not change the generality of the results.  Simulation parameters and methodology are described in the SI.
 \begin{figure}[htbp]
	\centering
	\includegraphics[width=\columnwidth]{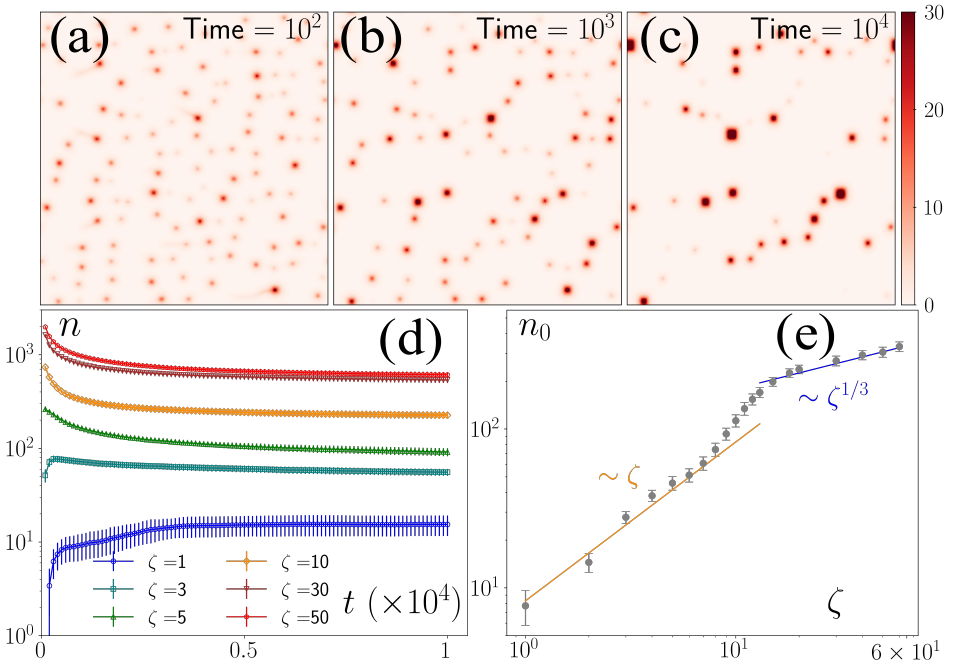}
  \caption{{\textbf{Defect coalescence dynamics}. (a-c) Coalescence dynamics of defects for $\zeta = 5$ at various stages. (d) Number of defects ($n$) over time ($t$) for different values of contractility, $\zeta$ (legend). (e) Steady-state defect count ($n_{0}$) exhibits a transition of $\zeta$ dependence from $\zeta^{1}$ to $\zeta^{1/3}$. Error bars are standard deviation of 109 replicates.}}
  \label{fig:aster_density}
\end{figure}
\paragraph*{Density of defects} To study the steady state behavior of the model, we numerically solved the equations for $10^7$ iterations ($dt=10^{-3}$) for different values of $\zeta$, keeping all other parameters fixed. We initialized the equations with $\mvec{p}(\mvec{r}) = (1,0)$ for all $\mvec{r}$ and used a disordered initial condition for $c(\mvec{r})$. The initial configuration rapidly evolved to a state with many defects, which then coalesced with each other to converge to a final state, where the number of defects, $n_0$, did not change appreciably over several million iterations (Fig.~\ref{fig:aster_density}a-d). $n_0$ is nonzero for all $\zeta$ values. However, the approach to $n_0$ is different above and below $\zeta = 5$ (Fig.~\ref{fig:aster_density}d-e). Below this threshold, the initial ordered state becomes linearly unstable through laminar structures at wave numbers $\mvec{q_d}$, whose magnitude $q_d$ scales as $\zeta^{0.5}$. For $\zeta > 5$, the ordered state transforms to a lattice of asters with each row alternating between inward (in-) and outward (out-) pointing asters. Because of the coupling of the $c$ field with the $\mvec{p}$ field, $c$ is depleted from the out-asters and enhanced at the in-asters (SI movies). Hence, local peaks in $c$ field correspond to the location of the in-asters. After the initial transient, the nonlinear effects become important, which leads to the nucleation of asters followed by their coalescence for $\zeta\leq5$, leading to a nonmonotonic evolution of asters towards $n_0(\zeta)$. For $\zeta > 5$, the number of asters decays to $n_0(\zeta)$ through coalescence of smaller asters to larger asters. The zeta-dependence of $n_0$ is nontrivial. For $\zeta \leq 5$, $n_0\sim \zeta^1$, which agrees with the prediction from the linearized theory, because the nucleation of the defects happen along the laminar structures, implying the scaling $n_0 \sim q_d^{-2}$.  In contrast, for $\zeta \geq 20$, $n_0\sim \zeta^{1/3}$, implying that the minimum separation between the defects scales as $\zeta^{1/6}$ (SI). The effect of nonlinearity becomes important for $5\leq \zeta\leq 20$, where $n_0$ shows smooth transition between the two regimes. The difference in the steady states in the different regimes is readily observed in the defect arrangements, which organize along 1D channels for $\zeta \leq 5$ and in liquid-like 2D structures for $\zeta \geq 20$, showing intermediate arrangements in between (SI movies).  

\begin{figure}[htbp]
	\centering
	\includegraphics[width=\columnwidth]{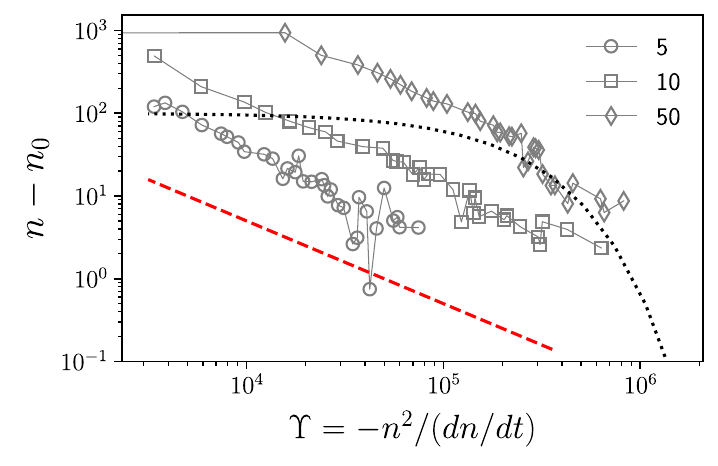}
	\caption{$n - n_0$ vs $\Upsilon$ for different $\zeta$ (legend); $n_0$ is the steady state defect number. In our model $n - n_0\sim \Upsilon^{-1}$ (red dashed line) and not exponential (black dotted line) as predicted by equilibrium kinetics~\cite{yurke1993coarsening}.}
	\label{fig:kinetics_comparison}
\end{figure}

\paragraph*{Coalescence kinetics}
 The dynamics of defects is fundamentally different from its equilibrium counterpart. In equilibrium, at zero temperature, all defects are annealed out to lower the free energy~\cite{CL-book}. The annealing kinetics is slow and shows logarithmic relaxation~\cite{yurke1993coarsening}. For example, below $T_{KT}$, the defect density, $n(t)$, scales as $n(t)\ln n(t)\sim t^{-1}$, which goes to zero as $t\rightarrow \infty$~\cite{yurke1993coarsening}. The relaxation kinetics of $n(t)$ in our model is clearly different from thermal equilibrium. To make the difference in kinetics evident, we compare how $n(t)$ varies with $\Upsilon(t) = -(n^{-2} dn/dt)^{-1}$~\cite{yurke1993coarsening}. When the equilibrium kinetics hold, $n(t)\sim e^{-A\Upsilon}$, where $A$ is a system-dependent constant. In contrast, in our model, we find that $n(t) -n_0 \sim \Upsilon^{-1}$, which confirms the fundamental difference between the active and the passive cases.

\begin{figure*}[htbp]
	\centering
	\includegraphics[width=2\columnwidth]{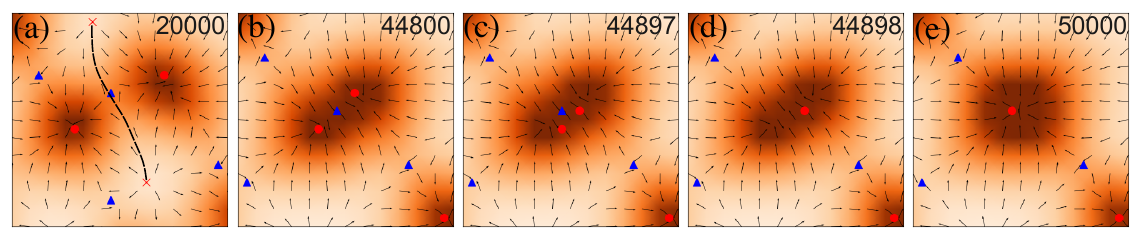}
	\caption{\textbf{Three-step defect coalescence}. (a) Pre-coalescent: shows defects far away from each other and the separatrix intact (black dashed line). (b) Step 1: As the defects come closer, the separatrix dissapears and their basin of attraction overlap.  (c-d) Step 2: First, two defects of opposite charges annihilate each other. (e) Step 3: the material is incorporated into the surviving defect, which relaxes to a symmetric shape. Number of iterations are shown in the panels. Red circle: in-asters (+1), red cross: out-asters (+1), and blue triangle: saddle (-1).}
	\label{fig:3-step}
\end{figure*}
\paragraph*{Three-step coalescence} To get a more microscopic understanding of the coarsening kinetics, we studied the coalescence of two in-asters. In between every two asters, there is a saddle point, which carries -1 topological charge (Fig~\ref{fig:3-step}). The saddle point connects two out-asters along a separatrix that separates the basin of attraction of the two in-asters (Fig~\ref{fig:3-step}a). The coalescence is facilitated through a large-scale reorganization of the $\mvec{p}$ field. In the immediate neighborhood of the two in-asters of interest, this reorganization results in the disappearance of the separatrix and the overlap of their basin of attractions (step 1, Fig~\ref{fig:3-step}b). After the overlap, one of the in-asters rapidly annihilate the saddle (step 2, Fig~\ref{fig:3-step}c-d), and the surrounding material relaxes to a rotationaly symmetric structure around the remaining defect (step 3,Fig~\ref{fig:3-step}e). What is surprising is that this process does not continue indefinitely. Once the defect density reaches $n_0$, the defects do not coalesce any more. Our initial hypothesis was that these are kinetically arrested states and would require longer simulations to continue the coalescence. However, even after running the simulation ten times longer than required to reach the steady state (Fig.~\ref{fig:aster_density}), we could not observe a single coalescence event. This observation is surprising because, in equilibrium, defects interact with each other through 2D coulomb interactions, which implies that at $T = 0$ the coalescence should continue until no defects are present. This observation, therefore, is consistent with the hypothesis that activity screens the effective interaction between the defects. The screening in the presence of activity is inherently different than that observed for $T > T_{KT}$ in equilibrium, where proliferation of many defects leads to the screening of the interaction and maintenance of the defect plasma. In contrast, here, the screening increases with $\zeta$, implying that the nonreciprocity of the active terms may be the origin of the screening, which has also been shown in the absence of number conservation~\footnote{Unpublished data from Pankaj Popli and Sriram Ramaswamy}. Understanding the active screening process requires the equation of motions of defects in 2D polar active materials, which are currently unavailable. Therefore, we resort to an indirect confirmation of this hypothesis by measuring the aster size distributions. 

\begin{figure}[htbp]
	\centering	\includegraphics[width=\columnwidth]{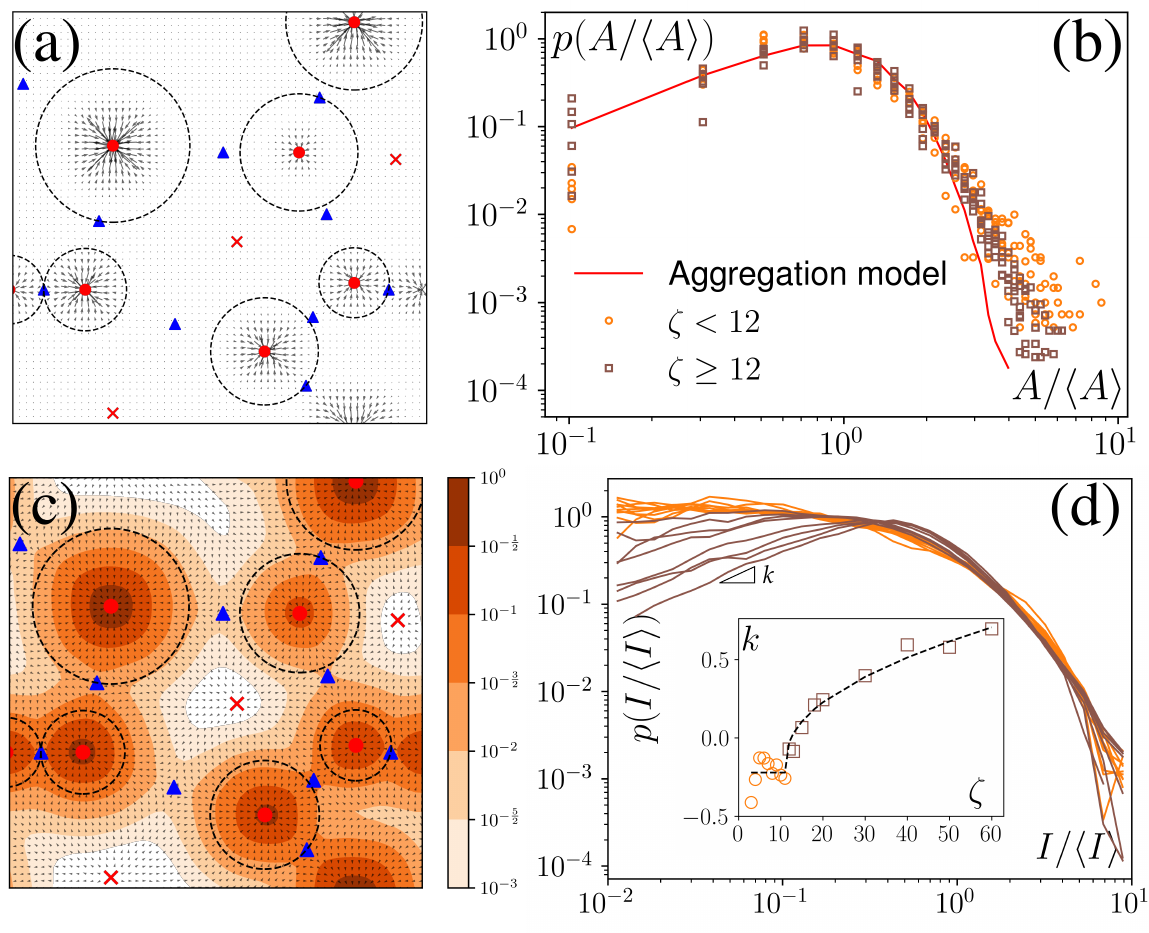}
	\caption{(a) The black dashed circles denote the region that defines the aster size, $A$.(b) Aster size distribution. Markers: Simulation data, solid line: aggregation model. (c) Aster intensity, $I$, is calculated by integrating the $c(\mvec{r})$ in the disk used to calculate the aster size. The normalized concentration levels are shown in the colorbar. (d) Probability distribution of $x = I/\langle I \rangle$ for different values of $\zeta$. For $x < 1$, $p(x) \sim x^{k(\zeta)}$ . Inset: $k$ vs $\zeta$. Black dashed line represents $(\zeta - \zeta_0)^{\beta}$, where $\beta = \frac{1}{2}$ for $\zeta > \zeta_0 \approx 12$.}
	\label{fig:aster_size}
\end{figure}

\paragraph*{Aster size distribution}  We have defined the size of the basin of attraction (BoA) of an aster as the size of the aster. However, true BoA is hard to calculate for our system and a simpler approximation must be found to generalize across the different system sizes and the parameters that we explore here. To this end, we have found that a disk centered on the aster core, whose radius is the distance to the nearest saddle,  is a good approximation of the BoA and is our measure of the size, $A$, of the aster (Fig.~\ref{fig:aster_size}a). The aster size distribution, $p(A)$ shows lognormal tails with power law head (Fig.~\ref{fig:aster_size}b). We can quantitatively reproduce $p(A)$ from a model of aggregation which only assumes that the aster interactions are screened beyond a cutoff distance $R_c$ and that the asters are immobile (SI). An aster increases its size by coalescing with the nearest aster until it exhausts all asters within $R_c$ or gets absorbed into another aster. For a wide range of $R_c$, $p(A)$ obtained from the simulation and the model are identical except at the tail (Fig.~\ref{fig:aster_size}b), which bolsters our hypothesis that the equilibrium long-ranged interactions between the asters is screened in the presence of activity.   

\paragraph*{Aster intensity distribution} In experiments, such as those with actomyosin complexes, identifying BoA is difficult, if not impossible. In contrast, the intensity of materials around the defects can be easily measured through optical microscopes, which is what we define here as the intensity of asters. Among all the defects considered here, the in-asters are the sinks, whereas the out-asters are the source. Hence, the $c$-field concentrates around the in-asters. Therefore, by integrating the $c$ field in the disk to measure $A$, we can calculate the intensity of the in-asters.   Because $c\mvec{p}$ vanishes exponentially outside the circle, the quantity of materials present outside the disk is negligible (Fig.~\ref{fig:aster_size}c). Hence, using our method, we can easily and accurately calculate the aster intensity for various parameters. The mean aster intensity $\langle I \rangle \sim n_0^{-1}$. Surprisingly, the aster intensity distribution, $p(I/\langle I\rangle)$, itself shows nontrivial transition with $\zeta$ (Fig.~\ref{fig:aster_size}d). For $I_{sc} = I/\langle I \rangle < 1$, $p(I_{sc}) \sim I_{sc}^{k(\zeta)}$. The power law exponent, $k(\zeta)$, interestingly, scales as $k \sim (\zeta - \zeta_0)^{1/2}$ for $\zeta > \zeta_0 \approx 12$, and as $(\zeta - \zeta_0)^0$ for $\zeta < \zeta_0$\footnote{The scaling of $k$ with $\zeta$ is suggestive and requires more careful investigation to ascertain the scaling exponent.}. The observed scaling indicates that the power law exponent is the order parameter of some underlying transition, which we cannot ascertain here. The value of $\zeta_0$ is also unclear due to noise in the data. A comprehensive investigation of this transition is beyond the scope of this manuscript.

\begin{figure}[htbp]
	\centering	\includegraphics[width=\columnwidth]{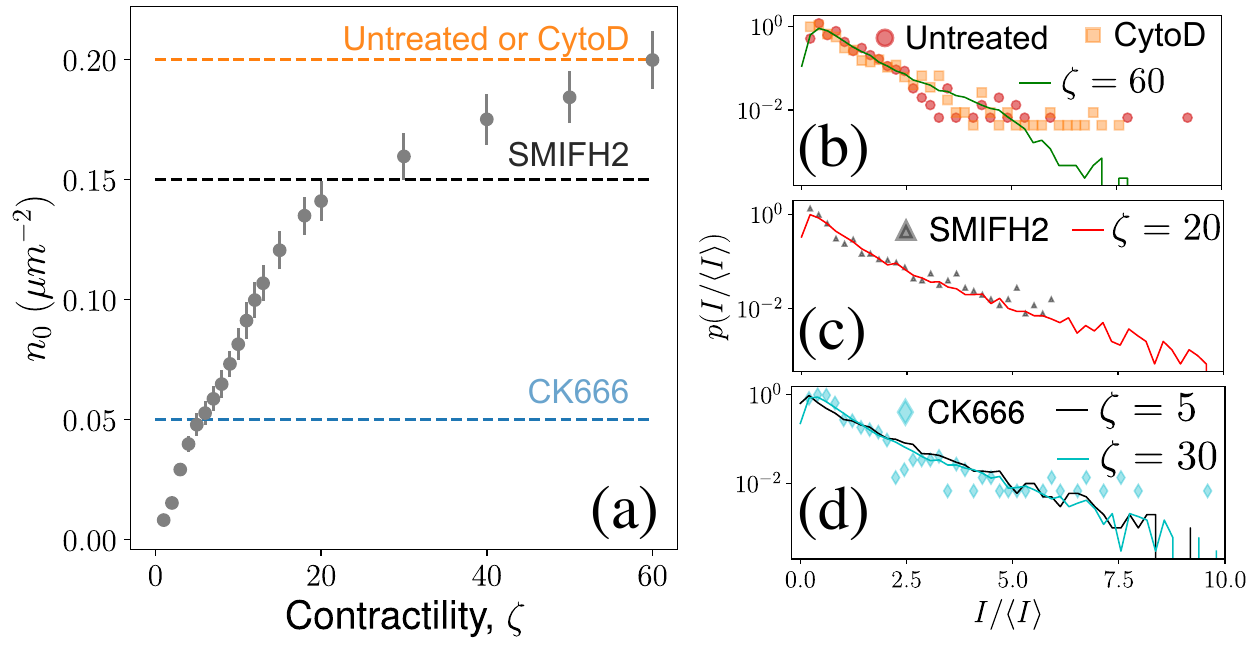}
	\caption{ \textbf{Experimental validation:} (a) Simulated in-aster density compared with experimental aster density under various drug treatments. (b-d) Comparison of aster area distribution from experiments~\cite{xia2019nanoscale} (markers) with $p(I/\langle I \rangle)$ distribution from simulations. Error-bar denotes standard deviation. }
	\label{fig:experiment}
\end{figure}

\paragraph*{Experimental evidence} Experiments on the actin cytoskeleton provides direct experimental confirmation of our results. For example, \textit{in vitro}
experiments on actomyosin complexes show coalescence of actin asters that are visually similar to our model: asters coalesce rapidly, the smaller asters get absorbed by larger asters, and the asters never coalesce into one single aster~\cite{soares2011active,wollrab2019polarity}. We expect similar dynamics to be present in live cells also. In fact, measurement of residence times of peripheral membrane proteins on the plasma membrane provides indirect evidence of the heterogeneous distribution of actin asters~\cite{sarkar2023lifetime}. To make this observation more concrete, we compared the steady state distribution of aster intensities from our model with that obtained from live cell experiments on mouse embryonic stem cells (mESCs)~\cite{xia2019nanoscale}. To make exact comparison, we converted the aster densities in our model to experimental units and compared it with the experimental aster densities (Fig.~\ref{fig:experiment}a). For untreated cells and for cells treated with the drug cytochalasin D (CytoD) the average density of the asters was approximately $0.2 \mu m^{-2}$. In our model, the aster density at $\zeta = 60$ is comparable to the above experiments, and we found that the aster intensity distribution for this $\zeta$ matches almost exactly with the experimental data (Fig.~\ref{fig:experiment}b). Treating the mESCs with the drug SMIFH2 reduces the aster density to $0.15 \mu m^{-2}$, which corresponds to $\zeta $ between 20 and 30 in our model. We found that $p(I)$ for  $\zeta = 20$ matches excellently with the experimental data (Fig.~\ref{fig:experiment}c), which predicts that SMIFH2 reduces the density of asters through the reduction of contractility. This observation is surprising because SMIFH2 is an inhibitor of Formin, a catalyst for actin polymerization, whereas contractility arises due to the effect of the motor protein myosin on the ACS. Interestingly, recent experiments suggest that SMIFH2 also inhibits myosin~\cite{nishimura2021formin}, which may explain our observation. Finally, we compared our model with CK666-treated cells, for which $\zeta = 5$ seemed the best match. However, the predicted $p(I)$ distribution did not match well with the experimental data for $\zeta = 5$, but matched well with the distribution for $\zeta = 30$ (Fig.~\ref{fig:experiment}d, see SI for the comparison). This mismatch suggested that CK666 lowers aster density not by lowering the contractility of the ACS. Indeed, CK666 is an inhibitor of the actin nucleating proteins Arp2/3, which has been shown to regulate the ACS architecture~\cite{xia2019nanoscale}. Therefore, our model needs to be modified appropriately to understand the effect of Arp2/3-dependent aster nucleation. 

From the experimental data~\cite{wollrab2019polarity}, it is evident that the asters never coalesce into a single aster even after many hours, which are orders of magnitude longer than the maximum timescales of our simulations. Therefore, it remains unclear whether the heterogeneous multi-aster state is a steady state or an arrested state. It is an important theoretical question, which will help us characterize the coarsening of topological defects in active polar system and contrast it with the equilibrium systems. Specifically, it was recently shown that a spontaneous proliferation of defects into multi-aster state destroys order in 2D active polar fluids without number conservation~\cite{besse2022metastability}. In that context, if the multi-aster state is an arrested state, we expect the order of the flocking state to be restored at long times. Whereas if the multi-aster state is a true steady state, then the ordered state is truly metastable. For our model, understanding this question will be important to understand the coarsening kinetics. In particular, we need to confirm whether the power law relaxation to the final state is a true relaxation to the steady state. The details are important and remains to be worked out in the future.  

Pragmatically, the distinction between an arrested metastable state and a steady state should be judged based on the relevant timescales of the problem. For a biological system, the important timescales are not the timescale to reach the final state, but the timescales of signaling, which are often much shorter than the time to relax to the final state. Similar considerations may apply even when we consider a material made out of active polar filaments, where the timescales of perturbations maybe much shorter than the relaxation timescales. As we have shown, there is excellent correspondence between experiments on actomyosin cortex and our model. Therefore, various predictions from our model, such as the phase transition in the intensity distribution, can be easily tested in existing experimental systems. The theory itself will be useful to understand general 2D active polar systems. Especially, the dynamics of multiple topological defects will prove to be useful to understand the mechanical and the rheological properties of these materials. These questions will be addressed elsewhere.

\begin{acknowledgements}
The authors would like to thank Pakorn Kanchanawong for sharing experimental data for Fig.~\ref{fig:experiment}. The authors would also like to thank Sriram Ramaswamy, Kripa Gowrishankar, Hugues Chat\'{e}, Anubhav A., and Saurav Varma for insightful discussions. SS thanks IISc, Axis Bank Center for Maths and Computing,  and SERB-DST India (SRG/2022/000163) for funding. SM thanks for the support received through the UGC-CSIR fellowship (211610108599) and PMRF fellowship (0203001). PP thanks IISc-IoE fellowship (80008199) for funding. 
\end{acknowledgements}
\bibliography{coarseningRefs}

\pagebreak
\widetext
\newpage
\setcounter{equation}{0}
\setcounter{figure}{0}
\setcounter{table}{0}
\setcounter{page}{1}
\makeatletter
\renewcommand{\theequation}{S\arabic{equation}}
\renewcommand{\thetable}{S\Roman{table}}
\renewcommand{\thefigure}{S\arabic{figure}}
\renewcommand{\bibnumfmt}[1]{[S#1]}

\begin{center}
\textbf{\large Supplementary Information: Coarsening of topological defects in 2D polar active matter}
\end{center}

  In these supplementary notes, we provide details on the hydrodynamic model used to describe the organization of the actin cytoskeleton in a cell. We discuss our numerical analysis along with the linear stability analysis of the coupled partial differential equations. We also discuss the static aggregation model used to rationalize the active screening of defect interaction. The notes are concluded with the details on experimental evidence of our findings.\\
\section{Model}
We investigate the coarsening of topological defects using a hydrodynamic description of polar active particles in 2D based on symmetry and conservation laws \cite{toner1998flocks,marchetti2013hydrodynamics}. This model is important for studying actin aster organization in the cell~\cite{gowrishankar2010active}. Actins are polar particles with head-tail asymmetry. Furthermore, myosin, a motor protein, binds to the actin filaments and drives them out of equilibrium by consuming ATP. The resultant active system forms various dynamic structures, including actin asters, which is a type of topological defect ~\cite{mermin1979topological}. We used the vector orientation field $\vec{p}(\vec{r},t)$ and the scalar concentration field $c(\vec{r},t)$ as hydrodynamic variables to describe this system.
\begin{eqnarray}
\cr    &&\frac{\partial \vec{p}}{\partial t} + \lambda(\vec{p}\cdot\vec{\nabla})\vec{p} = K\nabla^2\vec{p} + \zeta\vec{\nabla c} + \alpha \vec{p}-\beta |\vec{p}|^2 \vec{p} + \vec{f_p} \label{eq:SI-pfield}\\
\cr    &&\frac{\partial c}{\partial t}=-\vec{\nabla} \cdot (\vec{j_{d}} + \vec{j_{a}})= \vec{\nabla}\cdot(D\vec{\nabla}c-v_0c\vec{p})\label{eq:SI-cfield}
\end{eqnarray}\\
Eq.\ref{eq:SI-pfield} describes the time evolution of the orientational field. The terms $ K\nabla^{2}p ~\text{and} ~\alpha \vec{p}-\beta |\vec{p}|^2 \vec{p} $ of Eq.\ref{eq:SI-pfield} arise from the functional derivative (with respect to $\vec{p}$) of the free energy functional~\cite{marchetti2013hydrodynamics}, and $\vec{f_{p}}$ is the active noise in the vector field that comes due to stochasticity. The term $\lambda(\vec{p}\cdot\vec{\nabla})\vec{p}$ comes from the non linear advection. Symmetry requires that $\partial_{t} \vec{p}$ should contain a term proportional to $\vec{\nabla} c$, which contributes to aligning the polarization along gradients of the concentration field. 
The term $\left(\alpha \vec{p}-\beta |\vec{p}|^2 \vec{p}\right)$ comes due to the spontaneous polarization of the filaments \cite{gowrishankar2010active}.
Eq.\ref{eq:SI-cfield} describes the time evolution of the concentration field due to both advective currents, denoted as $\vec{j_{a}}$, and diffusive current, denoted as $\vec{j_{d}}$.  Due to the particle number conservation, the time evolution of the scalar field obeys the continuity equation. The fluctuation in the vector field is conservative and delta-correlated. Hence it obeys,
\begin{equation}
    \langle \vec{f_{p}}(\vec{r},t) \vec{f_{p}}(\vec{r'},t') \rangle = T_{a} \delta{(\vec{r}-\vec{r'})}\delta{(t-t')}
\end{equation}
where $T_{a}$ is the active temperature due to stochasticity and myosin binding-unbinding effect on the filaments. \\

Assuming a polar-ordered state with a uniform concentration field as $(p_{x},p_{y})=(1,0)$ and $c(\vec{r},t)=1$, we investigate the linear stability of this state to small perturbations to both vector and scalar field ($\vec{p} \rightarrow \vec{p}+\delta \vec{p}$, $c \rightarrow c+\delta c$). Linearizing and taking Fourier transformation ($\tilde{f}(\vec{k},t) = \int_{\vec{r}}  f(\vec{r},t) \, e^{i(\vec{k}\cdot\vec{r}-\omega t)} \, d^{2}\vec{r}$) of Eq.\ref{eq:SI-pfield} and \ref{eq:SI-cfield}, we get,\\
\begin{align}
    \partial_{t}
    \begin{bmatrix}
        \delta \tilde{c}(\vec{k},t) \\
        \delta \tilde{p_{x}}(\vec{k},t) \\
        \delta \tilde{p_{y}}(\vec{k},t)
    \end{bmatrix}
    =
    \mathbf{M}
    \begin{bmatrix}
        \delta \tilde{c}(\vec{k},t) \\
        \delta \tilde{p_{x}}(\vec{k},t) \\
        \delta \tilde{p_{y}}(\vec{k},t)
    \end{bmatrix}.
\end{align}\\
By examining the eigenvalues of the matrix $M$, we find the wave vector that makes the system unstable.
In the ordered state $\vec{n}(\vec{r},t)=0, n_{x}=1, n_{y}=0, c=1$, the matrix $M$ (taking $\alpha = \beta, \lambda=0$) is given by:\\
\begin{eqnarray*}
    \mathbf{M} =
    \begin{bmatrix}
        -Dk^{2} -ik_{x}v_0 & -ik_{x}v_0 & -ik_{y}v_{0} \\
        i\zeta k_{x} & (-Kk^{2} - 2\alpha) & 0 \\
        i\zeta k_{y} & 0 & -Kk^{2}\\
    \end{bmatrix}
\end{eqnarray*}

$\bullet$~{CASE-I, Pure splay term ($k_{x}=0$) } 
\begin{eqnarray}
\cr   \Lambda_{1} &=& 2\alpha - K k_{y}^{2} \\
\cr    \Lambda_{2/3} &=& \frac{1}{2}(-D-K)k_{y}^{2} \mp \sqrt{v_{0}\zeta} k_{y}
\end{eqnarray}

$\bullet$~{CASE-II, Pure bend term ($k_{y}=0$)}
\begin{eqnarray}
 \cr   \Lambda_{1} &=& - K k_{x}^2 \\
    \Lambda_{2} &=& -i v_{0 k_{x}} - \left(D-\frac{v_{0}\zeta}{2\alpha}k_{x}^{2}\right) + O(k_{x}^3) \\
 \cr   \Lambda_{3} &=& -2 \alpha - \left (K+\frac{v_{0}\zeta}{2 \alpha}\right) + O(k_{x}^{3})
\end{eqnarray}\\

The ordered state is stable without any contractility ($\zeta v_{0}=0$). The positive value of $\zeta$ introduces instability in the system characterized by a band of distortions centered around the splay distortion term $k_{x}=0$ and $k_{y}=q_{d}$ is given by,
\begin{eqnarray}
    q_{d} = 2\frac{\sqrt{\zeta v_{0}}}{(D+K)}
\end{eqnarray}

\section{Numerical Calculations}
\subsection{Simualtion of the hydrodynamic model}
We used explicit Euler FTCS (Forward in Time and Central in Space)~\cite{press2007numerical} integration method to solve Eq.\ref{eq:SI-pfield}  and \ref{eq:SI-cfield} numerically. We solved these two coupled non-linear partial differential equations (NPDEs) on   $L \times L$ system size with an iteration over $10^{7}$ steps. Where, $L=256$. We set the discretization in space $\Delta x=1$ and in time $\Delta t=10^{-3}$, which satisfies the Courant-Friedrichs-Lewy (CFL) condition and maintains the numerical stability criteria. In our simulation, we implemented periodic boundary conditions in both directions. We start from a uniformly ordered state with $(p_{x},p_{y})=(1,0)$ and $c= 1$. We add $c_{\sigma} = 0.001 \times \Gamma$, introducing a slight perturbation. Where $\Gamma$ is a random number chosen uniformly from the interval [0, 1]. In the simulation, we explore the behavior of this system by varying the contractility ($\zeta$) from 0 to 60 and keeping all other parameters fixed. For this simulation, we take $K$ and $D$ to be 2.5 and $\lambda, T_{a}$ to be zero, as taking a non-zero value of these parameters does not change the generality of the results~\cite{gowrishankar2010active, gowrishankar2016nonequilibrium}.

\subsection{Algorithm to detect topological defects}

\begin{figure}[h]
  \centering
 \includegraphics[scale=0.8]{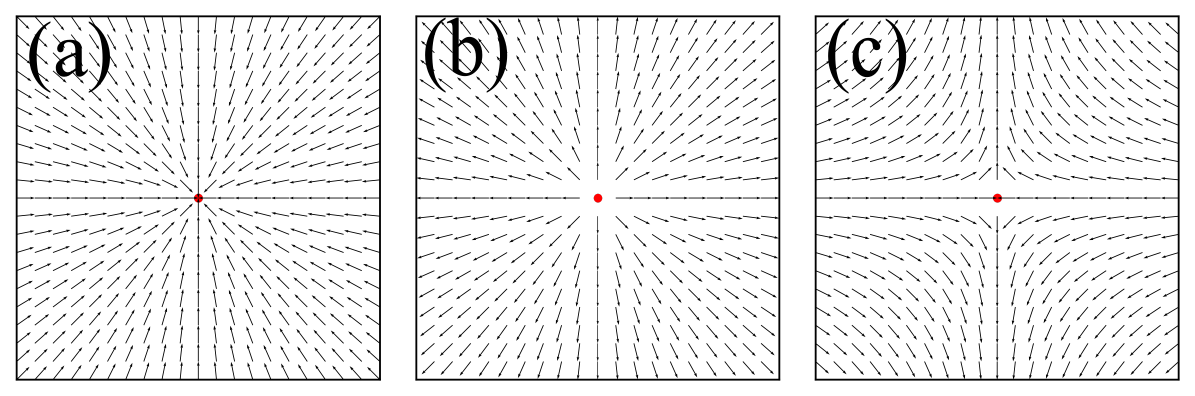}
    \caption{Topological defects with charges of +1 and -1, showcasing (a) positive inward, (b) positive outward, and (c) negative saddle configurations.}
    \label{fig:topodefect}
\end{figure}
Topological defects are singularities in the order parameter field that cannot be removed by a local perturbation in the field \cite{mermin1979topological}. The mathematical description of a topological defect is encapsulated by its winding number, which is given by,
\begin{eqnarray}
    S= \dfrac{1}{2 \pi}\oint_{c} \vec{\nabla}\theta \cdot \vec{dl} \label{windingnum}
\end{eqnarray}
  We utilize this mathematical framework to identify and locate defects in the system.
\begin{itemize}
    \item At each grid point $(i,j)$, we compute the values of $p_{x_{i,j}}$ and $p_{y_{i,j}}$ (x and y component of $\vec{p}$) and determine the corresponding angle $\theta_{i,j}$ using the inverse tangent function: $ \theta_{i,j} = \tan^{-1}\left(\frac{p_{y_{i,j}}}{p_{x_{i,j}}}\right)$.
    \item We then calculate the angle of the three nearest neighbors of the grid point $(i,j)$. To calculate the $\vec{\nabla} \theta$, we take the difference of their angles, which calculates $\mathcal I = 2 \pi S$.
    
    \item From $\mathcal I$, we calculate the winding number $S$ of the defect, which we identify as its topological charge ($\pm$ 1).
\end{itemize}

Depending on the configuration of the $\vec{p}$ field, we identify two basic types of topological defects with charges of $+1$ and $-1$ (Fig.~\ref{fig:topodefect}). The $+1$ charge corresponds to inward and outward pointing asters, while the $-1$ charge represents a saddle point. We use basic electrostatic principles to distinguish between inward and outward pointing +1 asters. By calculating the flux of the vector field around each +1 point and considering the sign of the flux, we can differentiate between these two configurations. Observing their positions, we integrate the concentration around these +1 inward defects to calculate the intensity of the asters.

\subsection{Parametrs:}
We characterize the length and timescale of the system by non-dimensionalizing the NPDE as,
\begin{eqnarray*}
   \cr l &=& \sqrt{D \tau} \\
   \cr  \tau &=& \dfrac{D}{v_{0}^{2}}
\end{eqnarray*}\\
In our simulation, we take $D = 2.5$ and $v_{0} = 1$, so the value of $\tau = 2.5 ~\text{and}~l = \sqrt{2.5 \times 2.5} = 2.5$. From the FCS studies \cite{goswami2008nanoclusters,gowrishankar2010active} of the living cell, the rotational diffusion constant is given by $D = 0.1 ~\mu m^2/sec$ ~\text{and}~ $ v_{0} = \sqrt{{D}/{\tau}} = 0.3~ \mu m/sec $. So, in the physical unit length, $l=\sqrt{0.1 \times 1}=0.31 ~\mu m$. The distance between two grid points $(dx=dy)$ is given by ${0.31}/{2.5} = 0.12 ~\mu m $. Also, for timescale, 1 second is equivalent to 2.5, so the physical time scale of each timestep ($dt$) is ${1}/{2.5} ~sec = 0.4~sec$. The total simulation time is $10^{-3} \times 10^{7} = 10^{4}$. So, the timescale in physical unit $10^{4} \times 0.4 ~sec ~\sim$ 1 hour and the length scale is $31 \times 31~\mu m^{2}$.   \\

\begin{center}
\begin{tabular}{ |p{4.2cm}|p{3cm}|p{3.5cm}|  }
 \hline
 \multicolumn{3}{|c|}{Parameter and Scaled Value} \\
 \hline
  Parameter (Dimensions) & Scaled Value & Physical Value \\
 \hline
  $K~(l^2/t)$ & 2.5 & 0.1 $\mu m^{2}/sec$\\
  $\zeta~(l^{3}/t))$ & $0 \rightarrow$ 60 & $0 \rightarrow 1.6 ~\mu m^{3}/sec$  \\
  $\alpha~(1/t)$ & 100 & 100 $sec^{-1}$\\
  $v_{0}~(l/t)$ & 1 & 0.3 $\mu m/sec$\\
  $D~(l^{2}/t)$ & 2.5 & 0.1 $\mu m^{2}/sec$ \\
 \hline
\end{tabular}
\end{center}\ \\
These values are chosen to closely align with previous experimental studies, ensuring that all terms are of the same order of magnitude \cite{goswami2008nanoclusters}.

\section{Size distributions and Experimental evidence}
The focus of our analysis is centered on the positive inward defects as the $c$-field concentrated only around the core of the in-asters. Calculating the size of the Basin of Attraction (BoA) involves finding the region around the core of an in-aster where the $p$-field flows towards the core. An approximate method to detect this region is described in the main text. The size obtained from this method is compared with the size obtained from a static aggregation model, which we describe here. Integrating the $c$-field in the BoA of each in-asters gives the intensity of the corresponding aster, which we compare with the experimental measurement of aster sizes. 

\subsection{Static aggregation model} 

The static aggregation model is inspired by the observation that the final state with heterogeneous aster sizes originate from an initial state where the asters (both inward and outward pointing) are arranged in a lattice. Because of their regular arrangement, the basin of attraction of each aster is exactly the same. Finally, we also observe that unless when they merge with each other, the asters do not move. When the asters merge, the timescale of merging is so small that it can be considered instantaneous. We distill these observations to construct the static aggregation model. In this model, 
\begin{enumerate}
	\item We initialize the system with asters placed on a lattice of lattice spacing $a$. All asters have equal size $A = 1$, i.e., the probability distribution function $P(x = A/\langle A\rangle) = \delta(x-1)$.
	\item We next perform Monte Carlo (MC) simulations until $P(x = A/\langle A\rangle)$ reached steady state. 
	\item The MC algorithm is as follows: 
	\begin{enumerate}
		\item Pick a random aster with $A\neq0$. Let's denote its area by $A_i$. 
		\item Identify all its neighbors within a radius $R_c$ with nonzero $A$.
		\item Find the nearest neighbor with $A\neq 0$. If there are multiple nearest neighbors, pick one randomly.  Let's call its area $A_j$.
		\item Merge asters $i$ and $j$ with the following criteria:
		\begin{itemize}
			\item If $A_i > A_j$, $j$ coalesces into $i$ and transfers its area to $i$, such that $A_j^{new} = 0$ and $A_i^{new} = A_i + Aj $. The opposite happens when $A_i < A_j$. This rule stems from the observation that the smaller asters coalesces with larger asters.
			
			\item If $A_i = A_j$, then pick any one of $i$ or $j$ with equal probability and merge it with the other.  
		\end{itemize}
		\item Stop simulation when the surviving asters have exhausted all asters within $R_c$ and $P(A/ \langle A \rangle)$ has reached a steady state.
	\end{enumerate}   
\end{enumerate}

\begin{figure}
	\centering
	\includegraphics{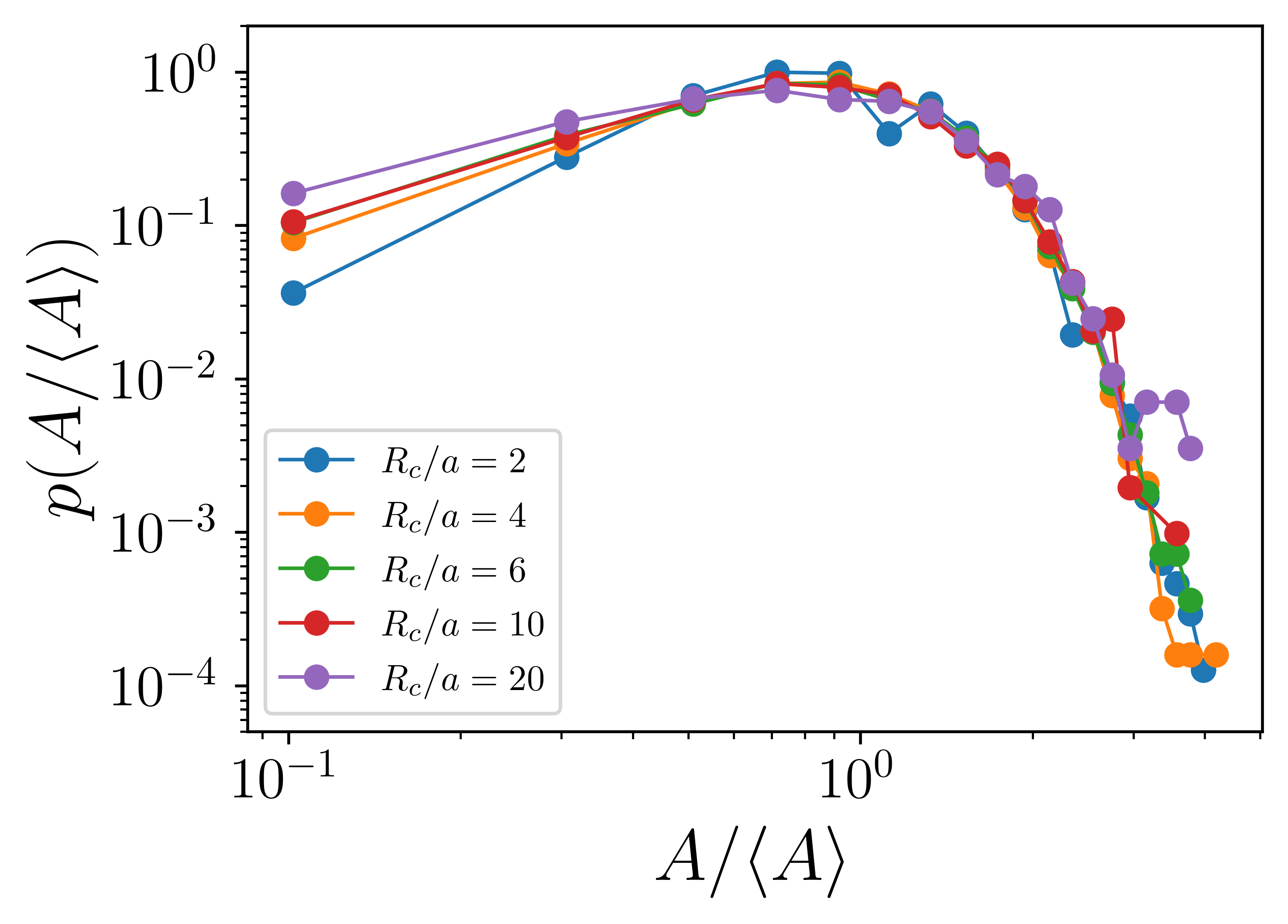}
	\caption{Probability distribution of scaled aster size $A/\langle A \rangle$ for different values of $R_c$.}
	\label{fig:aggregation_PDF}
\end{figure}

The evolution of the aggregation model from the initial state is shown in Fig.~\ref{fig:equilibrium_coarsening}. The steady state probability distribution is shown in Fig.~\ref{fig:aggregation_PDF}.
\begin{figure}
    \centering
    \includegraphics[width=0.5\linewidth]{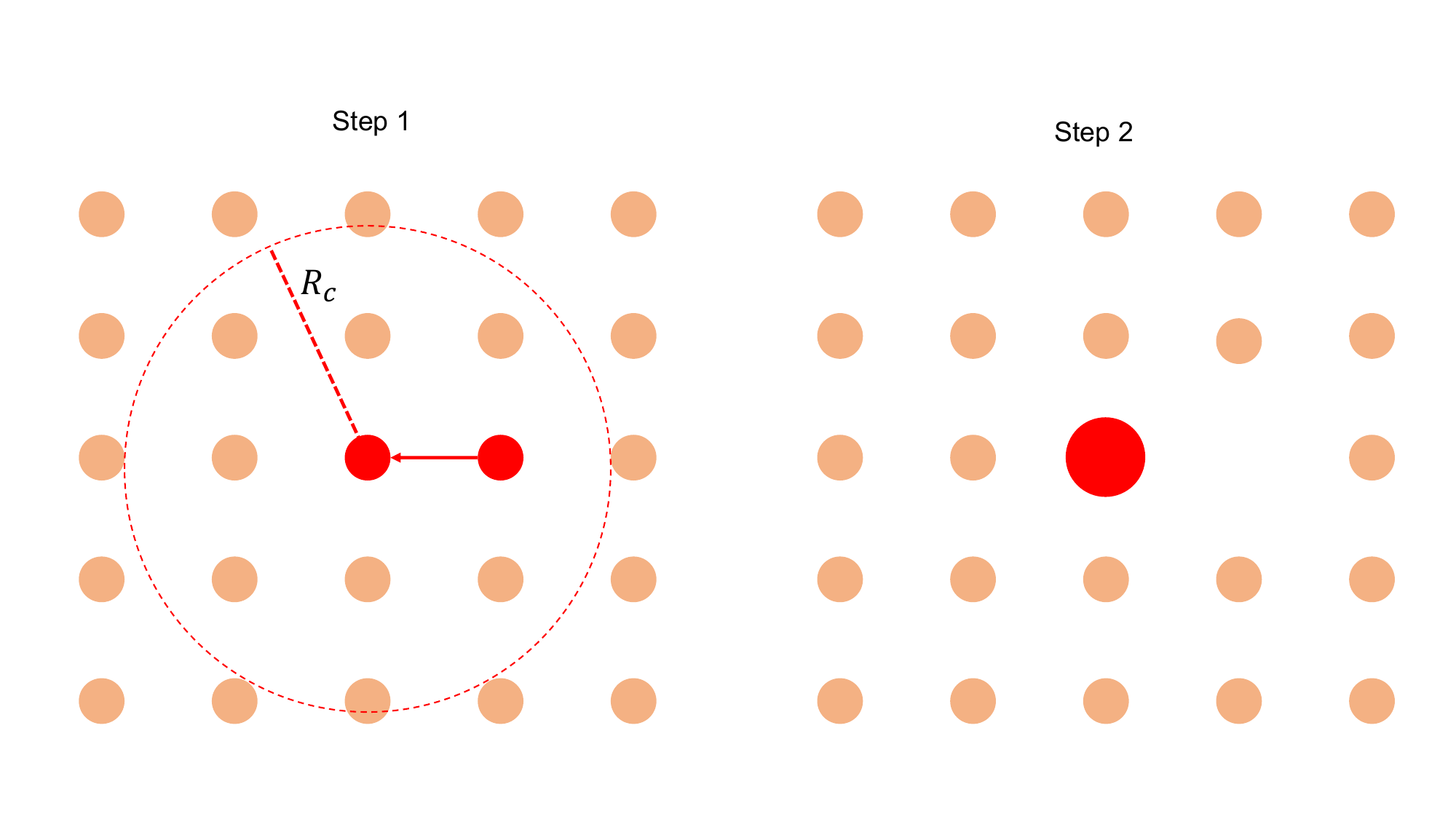}
    \caption{Figure shows the MC algorithm for merging process.}
    \label{fig:equilibrium_coarsening}
\end{figure}

\begin{figure}
    \centering
    \includegraphics[scale=1.5]{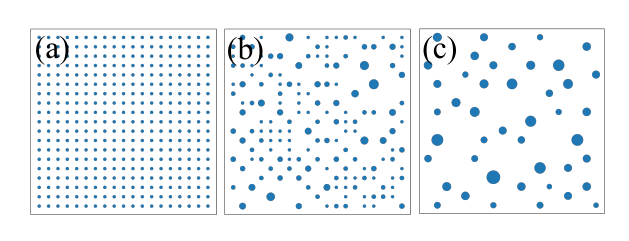}
    \caption{Time lapse images (a-c) show the coarsening process in the static aggregation model, where the materials can only merge when they are within a cutoff radius $R_{c}$.}
    \label{fig:equilibrium_coarsening}
\end{figure}

\section{Experimental Evidence}
We conducted a comparative analysis, aligning the distribution of aster intensities from our model with experimental data obtained from live cell experiments involving mouse embryonic stem cells (mESCs)\cite{xia2019nanoscale}. In the main text, we compared the probability distribution functions, which are sensitive to the choice of bins. Here, we compare the cumulative distribution functions (CDFs), which are free from this issue.  
\begin{figure}[H]
    \centering
    \includegraphics[scale=0.7]{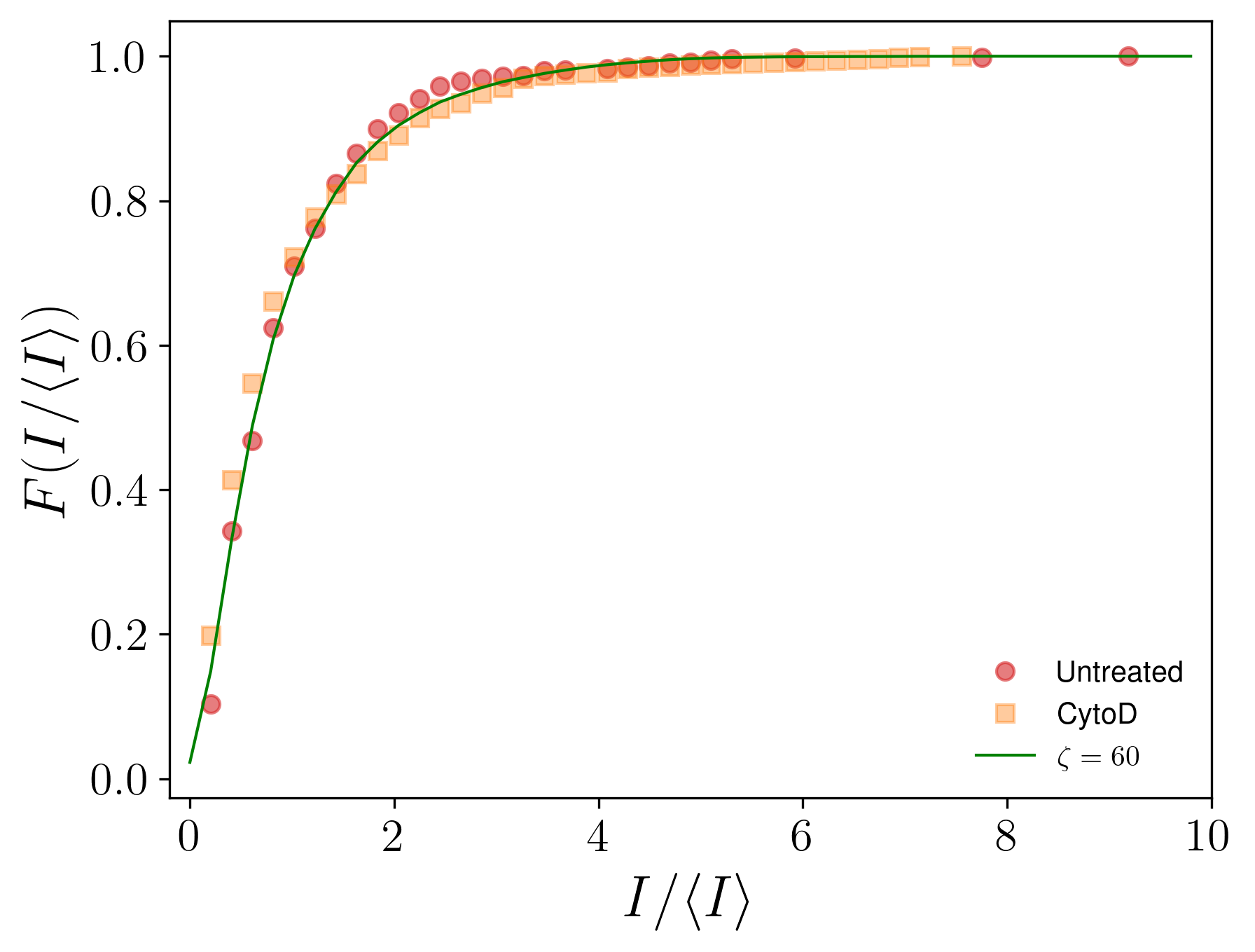}
    \caption{CDF comparing aster intensity distribution from experiments on Untreated cells and Cytochalasin D treated cells with the intensity distribution from our model.}
    \label{fig:CDF_Untreated}
\end{figure}

\begin{figure}[H]
	\centering
	\includegraphics[scale=0.7]{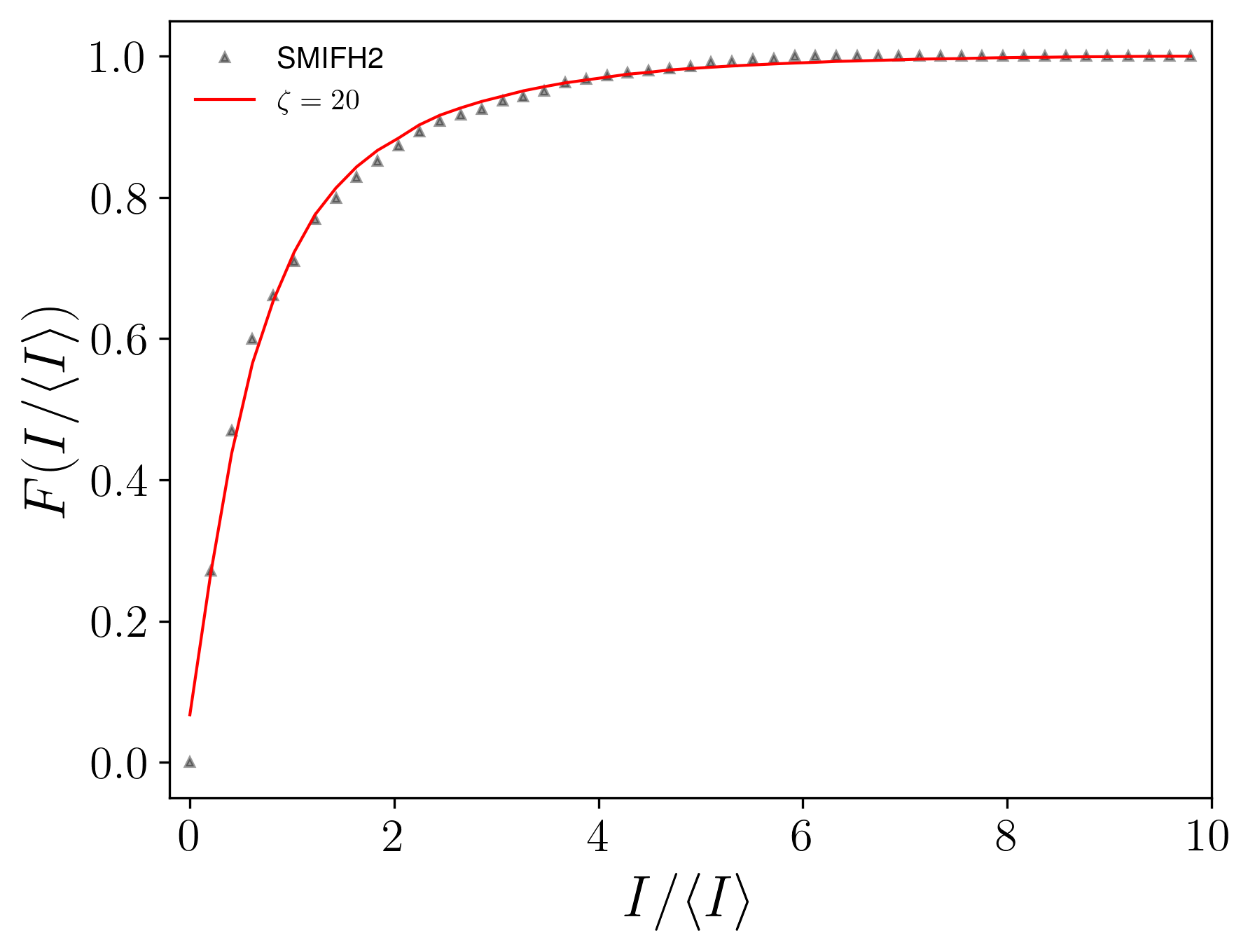}
	\caption{CDF comparing aster intensity distribution from experiments on SMIFH2-treated cells with the intensity distribution from our model.}
	\label{fig:CDF_SMIFH2}
\end{figure}

\begin{figure}[H]
	\centering
	\includegraphics[scale=0.7]{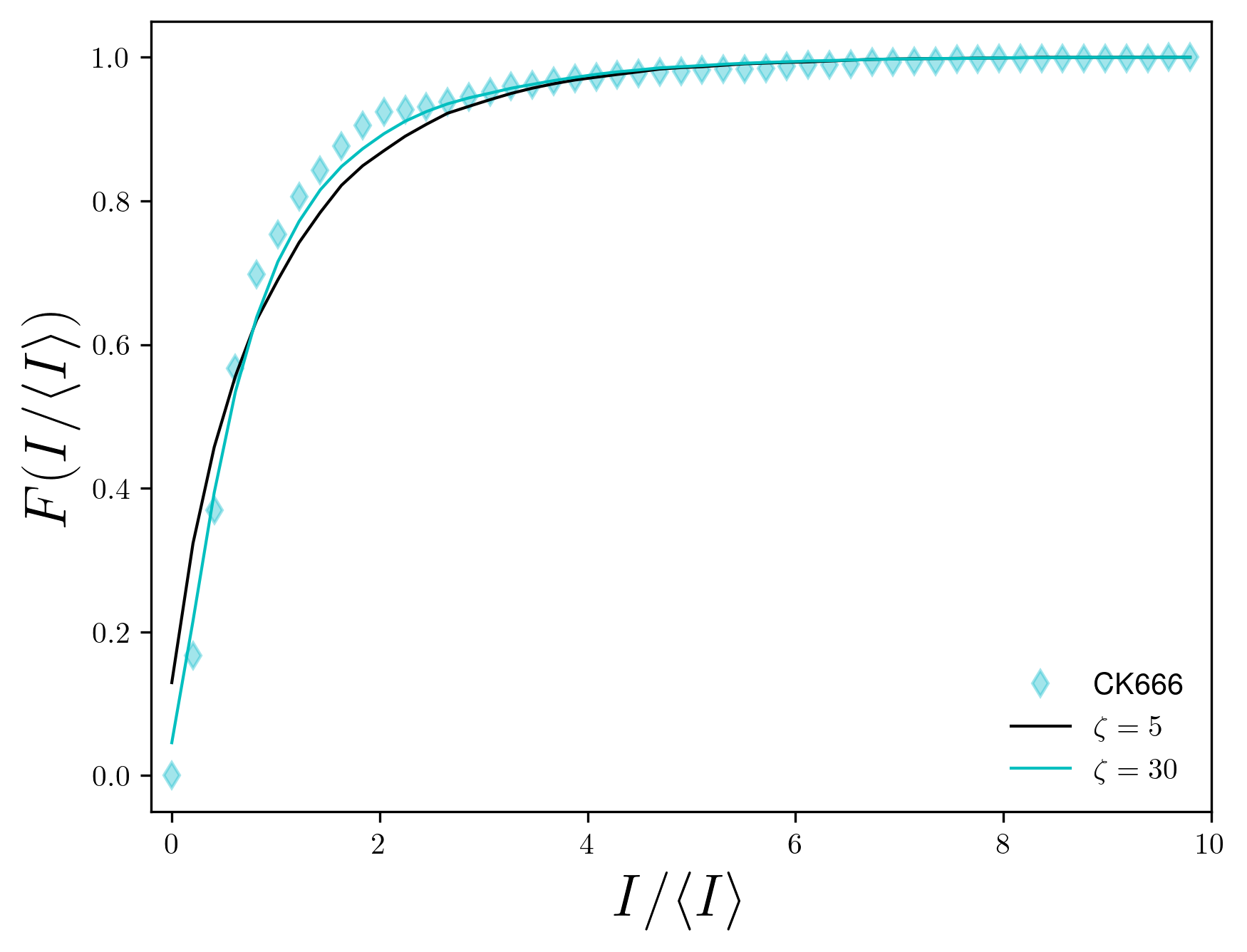}
	\caption{CDF comparing aster intensity distribution from experiments on CK666-treated cells with the intensity distribution from our model. Although our model predicts $\zeta = 5$ should correspond to the CK666 treated cells, in reality, $\zeta = 30$ matches experimental data better. This comparison suggests a potential limitation of our model.}
	\label{fig:CDF_SMIFH2}
\end{figure}

\section{Movie details} Three movies (\texttt{zeta\_1.mp4}, \texttt{zeta\_5.mp4}, and \texttt{zeta\_30.mp4}) are attached, which show the coalescence of actin asters, coarsening of topological defects with polar field and Schlieren texture for $\zeta = 1,5,30$ respectively.



\end{document}